\documentstyle[revtex,preprint]{aps}
\begin{document}
\def\rmts{RMT\ }
\def\rmt{RMT}
\def\vrhs{variable-range-hopping\ }
\def\goes{Gaussian orthogonal ensemble\ }
\def\gues{Gaussian unitary ensemble\ }
\def\gses{Gaussian symplectic ensemble\ }
\def\vrh{variable-range-hopping}
\def\goe{Gaussian orthogonal ensemble}
\def\gue{Gaussian unitary ensemble}
\def\gse{Gaussian symplectic ensemble}
\def\mit{metal-insulator transition}
\def\mits{metal-insulator transition\ }
\def\otimes{ {\textstyle \bigotimes} }
\def\trs{time-reversal\ }
\def\magnetoconductance{MC\ }
\def\quote#1#2#3#4{\squote {#1,\ {\sl#2}\ {\bf#3}, #4}.\par}
\def\qquote#1#2#3#4{\squote {#1,\ {\sl#2}\ {\bf#3}, #4};}
\def\nquote#1#2#3#4{\squote {#1,\ {\sl#2}\ {\bf#3}, #4}}
\def\book#1#2#3{\squote { #1,\ in {\sl#2}, edited by #3}.\par}
\def\nbook#1#2#3{\squote { #1,\ in {\sl#2}, edited by #3}}
\def\bbook#1#2#3{\squote { #1,\ in {\sl#2}, edited by #3}.}
\def\trans#1#2#3{[ {\sl #1} {\bf #2},\ #3 ]}
\def\prl{Phys. Rev. Lett.}
\def\pr{Phys. Rev.}
\def\ksi{\xi}
\def\porter{1}
\def\brody{2}
\def\metal{3}
\def\imry{4}
\def\altshuler{5}
\def\muttalib{6}
\def\lee{7}
\def\ucf{8}
\def\pichard{9}
\def\miller{10}
\def\efetov{11}
\def\error{12}
\def\kardar{13}
\def\nss{14}
\def\sivanb{15}
\def\experiment{16}
\def\schirmacher{17}
\def\us{18}
\def\rotor{19}
\def\friedel{20}
\def\halperin{21}
\def\shklovskii{22}
\def\landauer{23}
\def\cl{\centerline}
\hsize 15 truecm
\voffset 0.2 truecm
\hoffset -0.7 truecm
{\large \bf
\centerline{Random Matrix Theory of Transition Strengths and Universal}
\centerline{Magnetoconductance in the Strongly Localized Regime}}
\bigskip
\bigskip
\begin{tightenlines}
\cl{\Large Yigal Meir}
\medskip
\cl{Physics Department, University of California}
\cl{Santa Barbara, CA 93106}
\bigskip
\cl{\Large O. Entin-Wohlman}
\medskip
\cl{School of Physics and Astronomy}
\cl{Raymond and Beverly Sackler Faculty of Exact Sciences}
\cl{Tel Aviv University, Tel Aviv 69978, ISRAEL}
\bigskip
\bigskip
\noindent{
Random matrix theory of the transition strengths  is
applied to transport in the strongly localized regime. The crossover
distribution function between the different ensembles is derived and used to
predict quantitatively the {\sl universal} magnetoconductance curves in the
absence and in the presence of spin-orbit scattering. These
predictions are confirmed numerically.
}
\end{tightenlines}

\medskip

\bigskip

\newpage
Random matrix theory (RMT) has been used extensively and successfully in the
field of nuclear physics,$^{\porter,\brody}$ where
 the predictions of the theory concerning
 the distribution of excitation energies and the distribution
 of transition strengths agree well with
experimental data. The results of \rmts concerning the {\sl excitation-energy
distribution} have been successfully applied also
in condensed matter physics.$^{\metal-\muttalib}$
 In this work we demonstrate how \rmts
of {\sl transition strengths} can be applied in condensed matter theory, in
particular in the study of transport in the \vrhs regime. We obtain
{\sl universal} behavior of the magnetoconductance (MC) in the deeply localized
regime, which is determined by the crossover function between the
orthogonal to the unitary ensemble, in the absence of spin-orbit scattering,
and between the symplectic to the unitary ensemble in the presence of
strong spin-orbit scattering. We derive this crossover function,
and  predict quantitatively the universal \magnetoconductance curves,
which are verified numerically (Fig. 2).\par
The idea underlying the application of \rmts rests on the
 assumption that  the statistical behavior of  a complicated system
is determined by its symmetries. Accordingly, the Hamiltonian describing
the system can be classified into one of
three universality classes, characterized
by a parameter $\beta$ which counts the number of degrees of freedom
associated with each matrix element: The \goes ($\beta=1$), the \gues
($\beta=2$), where \trs symmetry is  broken, e.g. by a magnetic
field,  and the \gses ($\beta=4$), where rotational symmetry is
broken, e.g. by spin-orbit interactions. For these different ensembles one
can derive the distributions of level spacings and
 of overlap
probabilities, which in the case of nuclear transitions are related to the
distributions of excitation energies and  of transition
strengths, respectively.\par
The use of \rmts in condensed matter problems has been confined so far
to predictions concerning the level distribution. It was used for studying
electronic properties of small
metallic particles,$^{\metal}$ and it
 has been  also invoked$^{\imry,\altshuler}$
to explain the  theoretically predicted$^{\lee}$
universal conductance fluctuations in the weakly localized regime
 in terms of the rigidity of the level spectrum of the
transfer matrix$^\imry$ or the Hamiltonian.$^\altshuler$ Altshuler and
 Shklovskii$^{\altshuler}$ showed that when the Zeeman splitting is
 neglected,  the amplitude of the
conductance fluctuations is
 determined  by a parameter $1/\chi$, defined by
$ \chi = {{4\beta}/{s^2}}$  ,
where $s$ is the degeneracy
 of each level. Thus $\chi =$ (a) 1, (b) 2, (c) 4, (d) 8 for the cases (a) both
\trs and rotational symmetries are conserved, (b) only \trs
symmetry is broken, (c) only rotational symmetry is broken and (d) both
symmetries are broken ($\beta=2$).
 In particular, a magnetic field which breaks time-reversal
 symmetry decreases the magnitude of the fluctuations,
 independently of the amount of spin-orbit scattering present. This result
was confirmed by experiments.$^{\ucf}$\par
More recently Pichard et al.$^{\pichard}$ claimed that
similar arguments can also be applied to the strongly
 localized
regime. Here the conductance is determined by an equivalent
 resistor network$^{\miller}$ in which each two impurities are connected
by a conductance
 $g_0 J(H) \exp\left[-r_{ij}/\ksi(H)-\Delta\epsilon/kT\right]$,
where $g_0$ has units of conductance, $r_{ij}$ is the distance
between the impurities, $\Delta\epsilon=(|\epsilon_i|+|\epsilon_j|+
|\epsilon_i-\epsilon_j|)/2$, where $\epsilon_i$ are the energies of
the impurity states, and $\ksi$ is the localization length.
 Both $\ksi$ and the amplitude $J$ depend upon the
magnetic field.
 Using  RMT predictions for the eigenvalue statistics,
Pichard et al. concluded that $\ksi$ is doubled
 as \trs symmetry is broken and increases by a factor of four as
rotational symmetry is broken, in agreement with earlier exact
 results$^{\efetov}$
in quasi-$1d$ samples.$^{\error,\kardar}$ However, a relatively large
 magnetic field ($\equiv H_\ksi$)  of a unit quantum
flux through an area $\ksi^2$ is necessary
to induce a change in the localization length. (In such magnetic
fields other effects, such as the shrinking of the wave functions may be
significant, especially in doped semiconductors, where $\ksi$
 is on the order of the Bohr radius of the impurity state.)
On the other hand the amplitude $J(H)$ of the relevant hops is determined by
the interference of all paths within a cigar shaped area of length $R$ and
 width $\sqrt{R\ksi}$.$^{\nss,\sivanb}$ Accordingly, the relevant magnetic
field scale for a change in the amplitude, $H_R$,
is a unit flux through  the much larger cigar shaped area.
(Experimental values for the ratio between these two areas,
 $(R/\ksi)^{3/2}$,
 range from 5 to 100 and more, depending on temperature$^{\experiment}$).
Consequently,
the \magnetoconductance is
determined by the amplitude, for a wide range of magnetic fields.
Moreover, in the presence of strong spin-orbit scattering the localization
length is unaffected by a magnetic field,$^{\efetov,\kardar}$ and
the \magnetoconductance is dominated by the magnetic field
dependence of the amplitude. This amplitude is determined by the overlap
between the impurity wavefunctions. The overlap distribution
 has been calculated using various numerical and
analytic approximations.$^{\nss,\sivanb,\schirmacher,\us}$\par
In this work we use the fact that the
overlap probability between two wavefunctions
can be written in a form analogous to the transition
 strength in
nuclear physics,
 to  demonstrate how
the overlap distribution can be obtained similarly to the calculation
of the transition-strength distribution, using \rmt.
 Thus one expects this
distribution to be of a universal nature, leading to universal predictions
for the \magnetoconductance
in the \vrhs regime.\par
In order to derive the crossover distribution we study a general interpolating
$N\times N$ quaternionic matrix {\bf M},
\begin{eqnarray}
 {\bf M} = && \sqrt{1-{\gamma^2\over2}} \left[\sqrt{1-{3\delta^2\over4}}
{\bf S_0}\otimes {\bf I}
 + i{\delta\over2} \sum_{i=1}^3 {\bf A_i} \otimes {\bf \sigma_i} \right ]
\nonumber\\
 && \,\,\,+ \,\,i{\gamma\over\sqrt{2}} \left[\sqrt{1-{3\delta^2\over4}}
{\bf A_0}\otimes {\bf I}
 + i{\delta\over 2} \sum_{i=1}^3 {\bf S_i} \otimes {\bf \sigma_i} \right]  ,
\label{M}
\end{eqnarray}
where ${\bf S_i} ({\bf A_i})$ are symmetric (antisymmetric) $N\times N$ random
 matrices, whose elements are  normally distributed with a zero mean and unit
variance, and ${\bf \sigma_i}$ are the
Pauli matrices.
 For $\gamma=\delta=0$ the matrix $\bf M$ belongs to the \goe,
 for $\gamma=1$, $\delta=0$ it belongs to the \gue, and  for $\gamma=0$,
 $\delta=1$ it belongs to the \gse. Thus turning on
the parameter $\gamma$ from zero to $1$ interpolates between the Gaussian
orthogonal and the Gaussian unitary ensembles when $\delta=0$ and between
 the Gaussian symplectic and the Gaussian unitary ensembles
when $\delta=1$. Similarly changing
 $\delta$ from zero to $1$, with $\gamma=0$,  interpolates between the
Gaussian orthogonal and the Gaussian symplectic  ensembles.
 ${\bf M}$ is normalized
  such that the average overlap
is independent of the ensemble
(i.e. of $\gamma$ and $\delta$). The overlap probability
between sites $i$ and $j$, summed
over all final spin states and averaged over all initial spin states is
given by
\begin{eqnarray}
 y = & {1\over 2} {\em tr} \,{\bf M}_{ij}^{\dagger}{\bf M}_{ij}    =
(1-{\gamma^2\over 2})(1-{3\delta^2\over4})s_0^2+\nonumber\\
   &(1-{\gamma^2\over 2}){\delta^2\over4} {\displaystyle \sum_{k=1}^3a_k^2} +
  (1-{3\delta^2\over4}){\gamma^2\over 2}a_0^2+{\delta^2\over4}{\gamma^2\over 2}
  {\displaystyle \sum_{k=1}^3s_k^2} ,
\label{y}
\end{eqnarray}
where ${\bf M}_{ij}$ is the $2\times 2$ block of the matrix $\bf M$,
$s_k=({\bf S_k})_{ij}$ and $a_k=({\bf A_i})_{ij}$. The distribution of the
 overlap
probability, $P(y)$, is readily expressed in terms of its Laplace transform
$F(s)$,
\begin{eqnarray}
F(s)&  = \ \ \ {\displaystyle \int_0^\infty} e^{-sy} P(y)\ \ \ \ \ \ \ \ \ \ \
\
 \ \ \ \ \ \ \ \ \ \ \ \ \ \ \ \ \ \ \ \ \ \ \ \ \ \ \ \ \ \ \ \ \ \ \ \ \ \
\ \ \ \ \ \ \ \ \ \ \ \ \ \ \ \ \ \ \ \  \\
 = & 1/\left\{ [1+2(1-{\gamma^2\over 2}) (1-{3\delta^2\over4})s]^{{1\over2}}
 [1+\gamma^2(1-{3\delta^2\over4})s]^{{1\over2}}
 [1+{\delta^2\over 2}(1-{\gamma^2\over 2})s]^{{3\over2}}
 [1+{{\gamma^2\delta^2}\over 4}s]^{{3\over2}} \right\}.\nonumber
\label{ps}
\end{eqnarray}
For the four pure symmetry cases discussed above, which
correspond to (a) $\gamma=\delta=0$, (b) $\gamma=1$, $\delta=0$,
(c) $\gamma=0$, $\delta=1$, and (d) $\gamma=\delta=1$, the
overlap distribution function is given by
\begin{equation}
 P(y) =  {{(\chi/2)^{\chi/2}}\over{\Gamma(\chi/2)}} y^{\chi/2-1} e^{-\chi y/2}.
\label{py}
\end{equation}
Eq. (\ref{py}) is the exact result for the transition-strength distribution for
large matrices belonging to the Gaussian orthogonal ($\chi=\beta=1$),
the Gaussian unitary ($\chi=\beta=2$) and the Gaussian symplectic
($\chi=\beta=4$) ensembles. However,
when both \trs and rotational symmetries are broken,
$P(y)$ is given by  Eq. (\ref{py}), with $\chi=8$, even though the
$2N\times2N$  matrix {\bf M}  [Eq. (\ref{M})] belongs to the \gue. The
reason, as in the weakly localized regime,$^\altshuler$ is that even when \trs
symmetry is broken, the broken rotational symmetry couples the two spin
directions, so the overlap amplitude involves a sum over the two spins,
leading to eight free variables [see  Eq. (\ref{y})].\par
The distribution function can be  expressed analytically for
the crossover from the \goes to the \gues ($\delta=0$, finite $\gamma$),
 $P(y) =   {{e^{-y/Z} I_0 (\sqrt{1-Z}y/Z)}/{\sqrt{Z}}}$ ,
with $Z=2\gamma^2(1-\gamma^2/2)$,
which is the exact result for the transition-strength crossover function
for this case, in the limit of large matrices.$^{\brody,\sivanb}$ This
function has been verified numerically for a kicked
 rotor.$^{\rotor}$
 Similarly, the crossover function from the \goes to the \gses is given
by
\begin{eqnarray}
P(y)=&&\left[(1-{3\delta^2\over4})\delta^6/4\right]^{-1/2}\,\,y\,\,
 \exp\left[-\frac{y(1-{\delta^2\over
2})}{\delta^2(1-{3\delta^2\over4})}\right]\nonumber\\
&&\left\{I_{0}\left[\frac{y(1-\delta^2)}{\delta^2(1-{3\delta^2\over4})}\right]-I_{1}
\left[\frac{y(1-\delta^2)}{\delta^2(1-{3\delta^2\over4})}\right]\right\},
\label{gse}
\end{eqnarray}
 while the breaking of \trs symmetry in the \gses
($\delta=1$) gives rise to the crossover function
\begin{equation}
P(y)= 4 \left[ y\
\frac{e^{-2y/(1-{\gamma^2\over2})}+e^{-4y/\gamma^2}}{(1-\gamma^2)^2}
-{{\gamma^2}\over 2} \left(1-{{\gamma^2}\over 2}\right)
 \frac{e^{-2y/(1-{\gamma^2\over2})}-e^{-4y/\gamma^2}}{(1-\gamma^2)^3}\right] .
\label{gue}
\end{equation}
For general $\gamma$ and/or $\delta$, $P(y)$
can be written as
a convolution of two functions.
The small-$y$ behavior, however, can be directly deduced from the Laplace
transform  $F(s)$ and will be
described by the power law corresponding to the lower symmetry ensemble,
crossing over to the higher symmetry ensemble(s) behavior(s)
at values of $y$ which depend upon the values of $\gamma$ and $\delta$.\par

The most important consequence of Eq. (\ref{py}) is that the power law
that describes the small-$y$ behavior, the region that contributes the most
in the strongly localized regime, increases with a magnetic field (i.e.
with $\gamma$),  independently of the
 amount of spin-orbit scattering (i.e. of $\delta$). This leads to a positive
 \magnetoconductance  in the presence
and in the absence of spin-orbit
scattering.$^{\kardar-\sivanb,\schirmacher,\us}$
We checked this prediction by calculating the distribution of the
spin-averaged
transmission probability $T$,  at energy $E/V=0.1$,  through a $5\times5$
diamond,
described by the Anderson Hamiltonian of band width $4V$,
 with on-site uniform disorder
of width $W$.  We  include an orbital  magnetic field, characterized by
the overall flux through the diamond $\phi$, in units of $\phi_0$, the
 quantum flux, and  spin-orbit
scattering, characterized by the typical angle of rotation per hop
in spin-space, $\lambda$.$^{\friedel}$ In Fig. 1
 we plot the resulting small-$T$ distributions
   for four cases (a) $\phi=0$, $\lambda=0$,
 (b) $\phi=5.5$, $\lambda=0$, (c) $\phi=0$, $\lambda=2\pi$ and (d)
$\phi=5.5$, $\lambda=2\pi$, corresponding to the four cases discussed
above. The two panels in Fig. 1 correspond to the choices (a) $W/V=2$  and
 (b) $W/V=4$. We also plot in each panel the four
 slopes
resulting from Eq. (\ref{py}). There is clearly a satisfactory agreement
between the predictions and the numerical results. In particular, there
is a clear increase in the slope upon application of a magnetic
field. As the localization length only determines the average transmission
probability, the small-$y$ behavior is insensitive to the value of
disorder.\par
In order to calculate the conductance one has to solve the percolation problem
of the random resistor network,$^{\halperin,\sivanb}$
 with resistors distributed according to $P(y)$. This procedure  leads
to positive or negative \magnetoconductance for a system close to the
\mit, depending on the system parameters.$^{\sivanb}$
 However, away from the \mit, the
conductance of the system is given$^{\nss,\us,\shklovskii}$
 by $G = G_0 e^{<ln(y)>}$, except for an exponentially small region of
small magnetic fields. $G_0$ denotes a typical conductance of a single hop,
while $<...>$ denotes an average over the distribution $P(y)$.
Thus the relative MC, $\delta G = \left[G(H)-G(H=0)\right]
/G(H=0)$,
 will be {\sl universal} ---
depending only on the change in $P(y)$ as \trs symmetry is broken,
namely as $\gamma$ increases from zero.
In particular, this crossover function is given for the case of no
spin-orbit scattering ($\delta=0$) by $\delta G = \gamma\sqrt{2-\gamma^2}$.
Since we expect the \trs breaking parameter $\gamma$ to be proportional
to magnetic field, the \magnetoconductance is linear
at small fields,$^{\nss,\schirmacher,\us}$ saturating at a value
  of 1 for a strong enough
magnetic field.$^{\sivanb}$
 For strong spin-orbit scattering ($\delta=1$) the relative \magnetoconductance
is given by
\begin{equation}
\delta G = (1-{\gamma^2\over2})\left({{2-\gamma^2}\over\gamma^2}\right)
^{{3\gamma^4-2\gamma^6}\over{4(\gamma^2-1)^3}}
 \exp \left[{{\gamma^2-\gamma^4/2}\over{(\gamma^2-1)^2}}\right]  -  1.
\label{mc}
\end{equation}
At low fields the \magnetoconductance is now quadratic, saturating at
a value of $\exp(5/6)/2-1\simeq0.15$ at high fields. \par
We checked numerically these universal behaviors by calculating the
\magnetoconductance through the same $5\times5$ diamond giving
rise to Fig. 1. The conductance for a single realization was
 calculated from the transmission
probability, using Landauer formula,$^{\landauer}$ and was logarithmically
averaged over 100000 or 200000 realizations for
$W/V=25$, without spin-orbit scattering (Fig. 2a) and with strong
spin-orbit scattering (Fig. 2b). The high disorder was chosen so
that the localization length will be smaller than the size of the
sample, giving rise to the separation of the magnetic field
scale at which the localization length changes ($H_\ksi$)
 from the magnetic
field scale at which we expect our predictions to hold ($H_R$).
 As is
clearly seen in Fig. 2, there is a range of magnetic fields,
 corresponding approximately to one unit flux through the
sample,   where
the MC, as expected,  saturates. At higher fields,
the \magnetoconductance increases further as the localization
length starts to increase significantly. This effect is less
important in the presence of spin-orbit scattering because here,
for large enough samples, the localization length remains
unchanged by the magnetic field. We also plot in the figure the
prediction of our theory, in excellent agreement with the
numerical data. The only fitting parameter in both panels was the ratio
between the \trs symmetry breaking parameter, $\gamma$, and the
magnetic flux, $\phi$, which was chosen as $\gamma = 1.5 \phi$.
The shape of the curves is very similar to what one expects in experiments:
the \magnetoconductance at small fields should follow our universal
predictions,
while at higher fields other effects, such as the change in the localization
length and the shrinking of the wavefunction start to play a significant
role, and one expects  deviations from universality. The deeper in the
localized regime, the more separated the magnetic scales $H_R$ and $H_\ksi$
 become and the
more universal the \magnetoconductance should be.
\par
To conclude we have derived the crossover distribution for the overlap
probabilities (the transition strengths) between the different ensembles. This
distribution reduces to all the exact random-matrix-theory
 results in the appropriate limits.
This function has been used to predict the universal magnetoconductance
curves deep in the localized regime, where the localization length is
much smaller than the hopping length, and the percolation criterion
for the conductance is equivalent to the log averaging procedure.
In that case there is a wide range of magnetic fields, where
our universal predictions  hold. In this regime we expect all the
relative magnetoconductance curves, e.g. for different temperatures, to
 collapse onto a single curve, when expressed in terms of the scaled
\trs breaking parameter, presumably the magnetic flux through
the cigar shaped hopping area. So far experiments have been confined to
the vicinity of the \mit, since in this regime the conductivity is
more easily measurable. While one has to use more sensitive tools to probe
the deep localized regime, we hope that our work will stimulate further
experimental effort in this direction. \par

We thank B. Muhlschlegel, U. Sivan and N. S. Wingreen for helpful discussions.
Work at UCSB was supported by NSF
grant no. NSF-DMR90-01502
 and by the
NSF Science and Technology Center for Quantized Electronic Structures,
Grant no. DMR 91-20007.
This work was partially supported by the fund for basic research
administered by the Israel Academy of Sciences and Humanities and the
German-Israel Foundation for scientific research and development.
\newpage
\leftline{Figure Captions:}
{\noindent
\par
1. The distribution of the transmission probability through a $5\times 5$
diamond, described by an on-site disordered Anderson Hamiltonian, in
the presence of magnetic flux and spin-orbit scattering for two
values of disorder. The four curves correspond to (a) no magnetic field
and no spin-orbit scattering, (b) finite magnetic field (c) finite
spin-orbit scattering and (d) finite magnetic field and finite spin-orbit
scattering. Also depicted are the slopes (with arbitrary offsets)
 expected from Eq. (\ref{py}), (a) $-1/2$, (b) $0$, (c) $1$ and (d) $3$.\par
2. The magnetoconductance through a $5\times 5$ diamond for disorder
$W/V=25$. The solid lines are the predictions of the theory, where
the ratio $\gamma/\phi = 1.5$, between the \trs symmetry breaking
parameter $\gamma$ and the magnetic flux $\phi$, is the only fitting parameter.
\newpage
\begin{tightenlines}
\leftline{\sl references:}
\begin{enumerate}
\item C. E. Porter, {\em Statistical Theories of Spectra:
 Fluctuations} (Academic Press, New York, 1965); M. L. Mehta, {\em Random
Matrices} (Academic Press, New York, 1967).
\item T. A. Brody et al.,
 Rev. Mod. Phys. {\bf 53},
 385 (1981); J. B. French et al.,
 Ann. Phys. (NY), {\bf 181}, 198 (1988).
\item R. Kubo, J. Phys. Soc. Jpn {\bf 17},
975 (1962); L. P. Gorkov,
and G. M. Eliashberg, Zh. Eksp. Teor. Fiz. {\bf 48}, 1407 (1965)
 [Sov. Phys. JETP {\bf
21}, 940 (1965)]; R. Denton, B. Muhlschlegel, and D. J. Scalapino,
Phys. Rev. {\bf B7},
3589 (1973).
\item Y. Imry, Europhys. Lett. {\bf 1}, 249 (1986).
\item  B. L. Altshuler and B. I. Shklovskii, Zh. Eksp. Teor.
Fiz. {\bf 91}, 220, (1986) [Sov. Phys. JETP {\bf 64}, 127 (1986)].
\item K. Muttalib, J.-L. Pichard, and D. A. Stone, Phys. Rev. Lett. {\bf
59}, 2475 (1987); N. Zanon and J. L. Pichard, J. de Physique {\bf 49}, 907
(1988).
\item B. L. Altshuler, Pisma Zh. Eksp. Teor. Fiz. {\bf 41}, 530 (1985)
[JETP Lett., {\bf 41}, 648 (1985)]; P. A. Lee and D. A. Stone, Phys. Rev. Lett.
 {\bf 55}, 1622 (1985).
\item N. O. Birge et al., Phys. Rev. Lett.
 {\bf 62}, 195 (1989); D. Mailly et al.,
 Europhys. Lett. {\bf 8}, 471 (1989); P. Debray et al.,
Phys. Rev. Lett. {\bf 63}, 2264 (1989); O. Millo et al.,
Phys. Rev. Lett. {\bf 65}, 1494 (1990).
\item  J.-L. Pichard et al., Phys. Rev. Lett.  {\bf 65}, 1812 (1990).
\item A. Miller and E. Abrahams, Phys. Rev. {\bf 120}, 745 (1960).
\item K. B. Efetov and A. I. Larkin, Zh. Eksp. Teor. Fiz. {\bf 85}, 764 (1983)
 [Sov. Phys. JETP {\bf 58}, 444 (1983)].
\item It should be noted, however, that when \trs symmetry is broken in
 the presence of broken rotational symmetry, Pichard et al.$^{\pichard}$
 concluded that the
localization length is halved, in contradiction to the exact results of
Ref. \efetov\,  and the numerical calculations of Medina and Kardar (Ref.
\kardar)
and of this work.
\item E. Medina and
 M. Kardar, Phys. Rev. Lett. {\bf 66}, 3187 (1991);
 Phys. Rev. {\bf B 46}, 9984 (1992).
\item V. I. Nguyen, B. Z. Spivak, and B. I. Shklovskii, Pisma Zh. Eksp. Teor.
Fiz. {\bf 41}. 35 (1985) [JETP Lett. {\bf 41}, 42 (1985)]; Zh. Eksp. Teor. Fiz.
{\bf 89},
1770 (1985) [Sov. Phys. JETP {\bf 62}, 1021 (1985)].
\item U. Sivan, O. Entin-Wohlman, and Y. Imry, Phys. Rev. Lett. {\bf 60}, 1566
(1988); O. Entin-Wohlman, Y. Imry, and U. Sivan, Phys. Rev. {\bf B40}, 8342
(1989).
\item See e.g. O. Faran and Z. Ovadyahu, Phys. Rev. {\bf B 38}, 5457 (1988);
Q. Ye et al., Phys. Rev. {\bf B 41}, 8477 (1990).
\item W. Schirmacher, Phys. Rev. {\bf B 41}, 2461 (1990).
\item Y. Meir et al.,
 Phys. Rev. Lett. {\bf 66}, 1517 (1991).
\item K. Zyczkowski and G. Lenz, Z. Phys. {\bf B82}, 299 (1991).
\item J. Friedel, P. Lenglart, and G. Leman, J. Phys. Chem. Solids {\bf 25},
781 (1964). Y. Meir, Y. Gefen, and O. Entin-Wohlman, Phys. Rev. Lett. {\bf 63},
798
(1989).
\item V. Ambegaokar, B. I. Halperin, and J. S. Langer, Phys. Rev. {\bf B 4},
2612
(1971).
\item B. I. Shklovskii and B. Z. Spivak, in {\sl Hopping Transport in Solids},
edited by M. Pollak and B. I. Shklovskii (North Holland, Amsterdam, 1991).
\item R. Landauer, IBM J. Res. Dev. {\bf 1}, 233 (1957);
Phil. Mag. {\bf 21} 863 (1970);
M. B\"uttiker, \prl\  {\bf 57}, 1761 (1986).
\end{enumerate}
\end{tightenlines}
\end{document}